\documentclass[twocolumn,amsmath,amssymb,aps,longbibliography,floatfix]{revtex4-1}

\usepackage{cancel}
\usepackage{enumitem}
\usepackage[utf8x]{inputenc}
\usepackage{graphicx}
\usepackage{dcolumn}
\usepackage{bm}
\usepackage{hyperref}
\usepackage[switch]{lineno}
\usepackage{float}             
\usepackage{subfigure}
\usepackage{tikz}

\begin{document}

\title{Electronic spin-spin decoherence contribution in molecular qubits by quantum unitary spin dynamics}

\author{$^{1}$Alessandro Lunghi}
\email{lunghia@tcd.ie}
\author{$^{1}$Stefano Sanvito}

\affiliation{$^{1}$School of Physics, AMBER and CRANN, Trinity College, Dublin 2, Ireland}

\begin{abstract}
{\bf The realisation of quantum computers based on molecular electronic spins requires the design of qubits with 
very long coherence times, $T_{2}$. Dephasing can proceed over several different microscopic pathways, active 
at the same time and in different regimes. This makes the rationalisation of the dephasing process not straightforward. 
Here we present a computational methodology able to address spin decoherence processes for a general ensemble 
of spins. The method consists in the propagation of the unitary quantum spin dynamics on a reduced Hilbert space. 
Then we study the dependence of spin dephasing over the magnetic dilution for a crystal of Vanadyl-based molecular 
qubits. Our results show the importance of long-range electronic spin-spin interactions and their effect on the shape 
of the spin-echo signal.}
\end{abstract}

\maketitle
\section{Introduction}

In recent years, the possibility of using quantum properties of materials as active elements of new technologies has 
emerged, setting the foundation for a \textit{quantum revolution}. Quantum computing, even though theorised several years 
ago, has only recently been implemented in several physical systems, ranging from Cooper's pairs in 
superconductors~\cite{Nakamura1999} to electronic spins~\cite{Godfrin2017}. In a nut-shell, the vast dimension of a 
quantum system's Hilbert space and the quantum mechanical way to operate over it, allows one to perform operations 
at a speed much faster than in classical computers. The basic working principle of this technology requires the ability to 
initialise, manipulate and detect a many-body quantum state in its Hilbert space~\cite{DiVincenzo2000}. The coherence 
over time of the information ``stored'' in the quantum Hilbert space, generally referred as the characteristic time, $T_{2}$, 
is an essential property of a given system and depends on both intrinsic~\cite{Atzori2016b} and extrinsic factors~\cite{Zadrozny2015}. 
Quantum coherence has to be preserved long enough for the computation to be carried out and it represents the basic 
requirement for any class of potential qubits. 

Among the most promising paradigms to build quantum computers, transition-metal-based magnetic molecules have 
been brought forward as potential qubits, thanks to their long spin-coherence times and easy tunability\cite{Leuenberger2001,Ardavan2007,Winpenny2008,Lehmann2009,Wedge2012,Ferrando-Soria2016,Shiddiq2016,Gaita-Arino2019}. Indeed, molecular 
compounds represent a rather rich materials platform, where several synthetic strategies can be exploited to selectively 
tune specific inter- or intra-molecular interactions, with the goal of optimising the compound and extracting the basic 
physical picture of the dephasing. In recent years, an extensive amount of work has been performed in order to extend both 
the spin-lattice and spin-coherence times. Several guidelines to build more robust molecular qubits had been proposed, 
highlighting the importance of both the nature of the first coordination shell~\cite{Atzori2016b} and of surrounding nuclear 
spins density~\cite{Zadrozny2015}.

From a rational design point of view, the optimisation process has to be led by a detailed knowledge of those microscopic 
mechanisms leading to spin dephasing and this represents the focus of this work. Restricting our study to the sole spin-spin 
interactions, \text{i.e.} assuming the temperature to be low enough to exclude spin-lattice relaxation processes, several 
sources of spin dephasing might be operative. In a very general fashion, every spin in the system, other than the one addressed, 
cause a dynamical fluctuation of the spin local magnetic field, generating decoherence. In particular, three different types of 
spin-spin interactions are generally present in a crystal of molecular magnets: electronic-electronic dipolar interactions, nuclei 
hyperfine interaction and electronic-nuclear dipolar interactions. From a computational point of view, several approaches has 
been developed to study the central spin problem~\cite{Yao2006,Witzel2006,Yang2008,Maze2008}, where a single electronic 
spin loses its correlation due to the interaction with a bath of nuclear spins, finding application in both solid-state defect 
quantum bits~\cite{Seo2016} and molecular qubits~\cite{Bader2016a}. In order to complement the existing picture of spin 
dephasing, we present here a computational strategy to address the role of electronic spin-spin interactions over the dephasing 
time for an ensemble of generic spins. This approach, able to address up to hundreds of spins at the same time, is used to simulate 
the spin-echo decay experiment for a crystal of Vanadyl qubits, revealing a linear dependence of $T_{2}$ with respect to the 
magnetic dilution. Moreover, insights regarding the shape of the spin-echo envelop are reported, showing a transition from a 
Gaussian profile, for a perfectly ordered crystal, to a stretched exponential, for the random-distributed magnetically diluted phase.

\section{Unitary Quantum Dynamics and Spin Echo}

A spin initialised in a specific state is always subject to dephasing, unless it represents a perfect close system, namely it is
isolated from the rest of the universe. Such a strict condition is never met in practice, so that a quantum system is never fully 
isolated and thus is subject to some non-trivial dynamics. The study of this process, in principles, requires the knowledge of 
all the spin-``external world'' interactions and the solution of the Schr\"odinger equation for the coupled system. 

In order to formalise the problem, let us introduce the total spin Hamiltonian for a system of $N_\mathrm{S}$ coupled spin 1/2,
$\mathbf{S}(i)$, 
\begin{equation}
\hat{H}_\mathrm{S}=\sum_{i}^{N_\mathrm{S}}\mu_\mathrm{B}\vec{\mathbf{B}}\cdot\mathbf{g}(i)\cdot\vec{\mathbf{S}}(i)+\frac{1}{2}\sum_{ij}^{N_\mathrm{S}}\vec{\mathbf{S}}(i)\cdot\mathbf{D}^\mathrm{dip}(ij)\cdot\vec{\mathbf{S}}(j)\:,
\label{SH}
\end{equation}
where the interaction of the spins with an external field, $\vec{\mathbf{B}}$, is gauged by the Bohr magneton, $\mu_\mathrm{B}$, 
and the $g$-tensor, $\mathbf{g}(i)$, while the spin-spin interaction, $\mathbf{D}^\mathrm{dip}$, is assumed to be only dipolar in 
nature. The state of the spin system can be described in term of the spin density matrix, $\rho$, whose dynamics is regulated by 
the Liouville equation
\begin{equation}
\frac{d\hat{\rho}}{dt}=-\frac{i}{\hbar}[\hat{H}_{s},\hat{\rho}]\:.
\label{UN}
\end{equation} 
The matrix elements of $\hat{H}_\mathrm{S}$ are usually computed on a basis set corresponding to the tensor product of all 
single spins $\hat{S}_{z}(i)$ operators' eigenkets
\begin{equation}
\{|S_{z}(1),S_{z}(2),...,S_{z}(N)\rangle\}=\bigotimes_{i=1}^{N}\{|S_{z}(i)\rangle\}\:.
\label{Szbasis}
\end{equation}
By diagonalising $H_\mathrm{S}$ one can solve numerically Eq.~(\ref{UN}) to obtain an expression for the time 
evolution of $\rho_{ab}$ in terms of the eigenvalues of $H_\mathrm{S}$, $E_{a}$,
\begin{equation}
\rho_{ab}(t)=U_{ab,ab}(t-t_{0})\rho_{ab}(t_{0})=e^{-i\omega_{ab}(t-t_{0})}\rho_{ab}(t_{0})\:,
\label{UN1}
\end{equation}
where $\omega_{ab}=(E_{a}-E_{b})/\hbar$ and $U_{ab,ab}(t-t_{0})$ is the unitary evolution operator.

Eq.~{\ref{UN1}} describes the exact dynamics of the $N_\mathrm{S}$ interacting spins, but its use is limited by 
the size of the matrix that one can actually diagonalise. The complexity of the problem arises mainly from the 
exponential dependence of the size of the spin Hilbert space on $N_\mathrm{S}$. This limits the solution to $\sim16$ 
spins on common parallel machines, a general problem in quantum many-body physics. Nevertheless, several 
approaches to overcome such limitation exist. The specific problem concerning the correlation loss of a central 
spin embedded in a spins bath has been successfully tackled with a range of methods falling under the name of 
\textit{cluster approximations}~\cite{Yao2006,Witzel2006,Yang2008,Maze2008}, where the total contribution to 
dephasing is decomposed into the contribution of small spin clusters. Another possible approach, somehow based 
on a similar physical principle, consists in working on a truncated Liouvillian space, where the states considered 
are chosen among the ones deemed relevant. The basic idea behind these methods consists in assuming that only 
a small part of the total Liouville space will be actually explored during the time evolution. Moreover, the states explored
will be those close enough to the states at time $t=0$. Coupled clusters~\cite{Roger1990}, density-matrix renormalisation 
group~\cite{Daley2004} and adaptive state-space restriction~\cite{Kuprov2007} approaches all fall into this category. 

In this paper we will take advantage of a non-perturbative Hilbert space size reduction technique, belonging to the 
last class of methods and inspired by Kuprov's works~\cite{Kuprov2007}. In the presence of an external magnetic field 
pointing along the $z$ direction and in the limit $B_{z}\rightarrow\infty$, the ground state will tend to a state with all the 
spins aligned towards the field. In this situation, the ground state can be represented with only one vector in the $S_{z}$ 
basis set and all the excited states are represented by single, double, ..., $N$-th fold spin flips. In the presence of spin-spin 
interactions a mixing between these eigenstates is introduced, generating a non-trivial dynamics, leading to decoherence.
However, as long as the Zeeman term remains the leading interaction, the nature of the eigenstates will be only slightly affected. 
Therefore, by assuming that the starting configuration can be generated close to the ground state of the system, we can 
restrict the size of the Hilbert space to those states within a few spin flips from the initial state. Such a restriction makes the 
total number of spin states needed to investigate the dynamics to scale polynomially with $N_\mathrm{S}$ and the problem 
expressed by Eq.~(\ref{UN}) and Eq.~(\ref{UN1}) becomes easily tractable even for $10^{2}$ spins. 

The method just outlined is here employed to study the effect of electronic spin-spin interactions in the decoherence 
processes through the simulation of the spin Echo experiment. The spin Echo experiment is often use as a mesure of T$_{2}$, as it averages out inhomogeneous effects, and it consists in a two pulses sequence. The first pulse brings the magnetization along the x-axis, then, after a free evolution time $\tau$, a $\pi$ pulse is applied. After the second pulse, another interval $\tau$ of free evolution time is waited before reading out the x component of the residual magnetization. This can be described by the path,
\begin{equation}
\rho(0)\xrightarrow{R(\frac{\pi}{2})}\rho^{\frac{\pi}{2}}(0)\xrightarrow{U(\tau)}\rho^{\frac{\pi}{2}}(\tau)\xrightarrow{R(\pi)}\rho^{\frac{3\pi}{2}}(\tau)\xrightarrow{U(\tau)}\rho^{\frac{3\pi}{2}}(2\tau)\:,
\label{echo}
\end{equation}
where $\hat{\rho}(0)$ is the starting density matrix, $U(t)$ is the time propagator operator as expressed in Eq.~(\ref{UN1}) 
and $R(\theta)$ is a rotation operator that simulates the effect of a micro-wave electron paramagnetic resonance (EPR) pulse.

\section{Computational Details}

The implementation of the method just outlined follows three steps: 1) the construction of the total Hamiltonian matrix, 
2) the diagonalization and 3) the propagation of an initial density matrix according to the scheme introduced in Eq.~(\ref{echo}). 
The truncated basis set is generated by flipping a given number of spins with respect to a state where they are all 
aligned along the field direction. When possible, the spin Hamiltonian is rotated in a block-diagonal form by forming 
a translation-invariant basis set. The spin Hamiltonian diagonalisation is performed with scalapack libraries~\cite{Choi1996}. 
The initial spin-density matrix is generated as a pure state where all the spins are aligned along the external magnetic 
field. Tests have also been performed with an initial spin-density matrix equal to the canonical ensemble distribution, but 
no differences in the extracted T$_{2}$ values have been observed. The rotation operator is constructed by converging 
a Taylor expansion of $R(\theta)=\mathrm{exp}[i(\vec{\mathbf{S}}\cdot\vec{\mathbf{n}})\theta/2\pi]$, where $\vec{\mathbf{S}}$ 
is the spin-vector operator of the spin to be rotated, $\vec{\mathbf{n}}$ is the normal rotation axis and $\theta$ is the angle of 
rotation. The time evolution of the expectation value for a general operator $\hat{A}$ is 
$\langle A(t) \rangle = \mathrm{Tr}\{\hat{A}\hat{\rho}(t)\}$. 

We have also tested an alternative to the diagonalisation of the spin Hamiltonian matrix that involves the use of the original 
$S_{z}$ basis set. In this basis set almost all the operators have a very sparse matrix representation that can be conveniently 
exploited for storing purposes. However, the time propagator $\hat{U}(t-t_{0})$, being an exponential operator, has a dense 
matrix representation and must be multiplied to the density matrix twice at each time-step. Moreover, the generation of the 
exponential operator in this basis set requires the use of a very short time step $\Delta t=t-t_{0}$, usually requiring the calculation 
of more than 10$^{6}$ steps to reach the desired microsecond time-scale. Consequently, here we employ the direct diagonalisation 
approach that makes it possible to set the desired time step in the dynamics propagation. The $\mathbf{g}$ operators, when 
not available from experiments, are computed by CASSCF calculations with the software {\sc Orca}~\cite{Neese2012}. 
An active space including the $d$ electrons and five 3$d$ orbitals is used. The spin-spin distance used to compute the dipole-dipole 
interaction is implemented using periodic boundary conditions in order to correctly reproduce the solid state environment. 

\section{Results}

In this work we study two molecular systems belonging to the Vanadyl class of molecular quibits: VO(acac)$_{2}$~\cite{Tesi2016}, 
being acac=acetylacetonate, and VO(dmit)$_{2}$~\cite{Atzori2016b}, with dmit = 1,3-dithiole-2-thione-4,5-dithiolate. These 
magnetic molecules have recently become subjects of intense research as they naturally show rather long coherence times, 
up to several $\mu s$ even at room temperature~\cite{Atzori2016}, suggesting the possibility of further performance increase 
if specifically tailored.  

\begin{figure}[h!]
\includegraphics[scale=1]{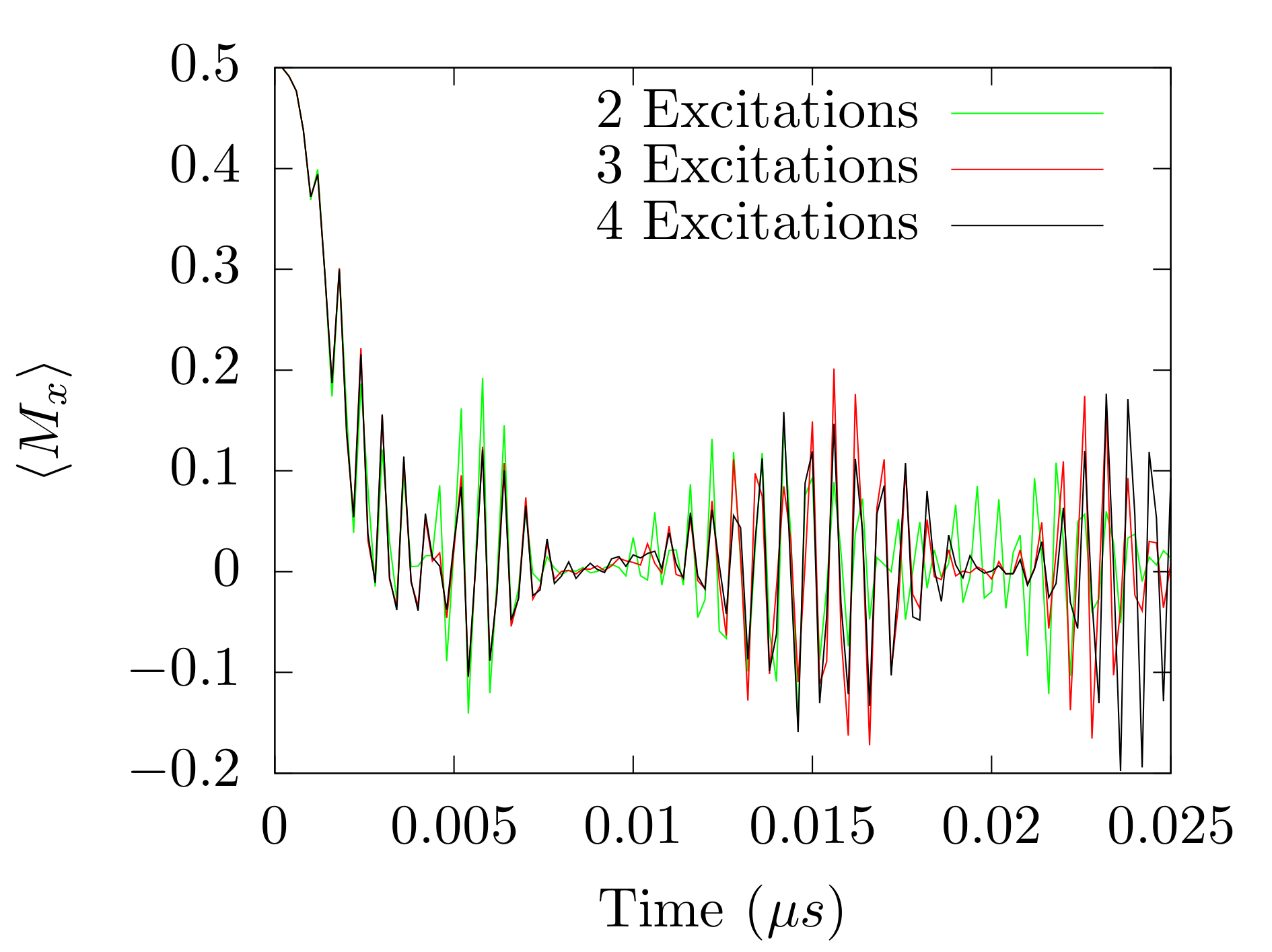}
\caption{\textbf{Spin Echo dependence on the number of spin excitations.} The spin-echo profile for the dynamics of 16 spins, 
arranged on a 2$\times$2$\times$2 supercell of the VO(acac)$_{2}$ crystal. The time evolution is computed when 2 (green curve), 
3 (red) and 4 (black) spin excitations are included in the basis set. Note that the minimal basis set, containing 2 spin excitations
only, is already sufficient to describe the short-time dynamics.}
\label{echonex}
\end{figure}

The first system investigated is the VO(acac)$_{2}$ crystal. The phase-coherence properties of this compound have not 
been investigated experimentally, and here we use it to show the consistency of our computational method. The $\mathbf{g}$ 
tensor for this specie is calculated as explained in the ``Computational Details'' section and it reproduces the typical values 
for this class of molecules: $g_{xx}=1.9430$, $g_{xy}=0.0179$, $g_{xz}=-0.0170$, $g_{yy}=1.9749$, $g_{yz}=0.0084$ and $g_{zz}=1.9735$.
The orientation of the molecule in space is as found in the crystal's unit-cell aligned with the crystallographic $\vec{\mathbf{a}}$ vector along $x$ and the $\vec{\mathbf{b}}$ vector lying in the $xy$ plane. The lattice parameters can be found in ref. \cite{Tesi2016}. Finally the external field was chosen aligned along the $z$ direction.
Fig.~\ref{echonex} reports the time evolution of the expectation value of the transverse magnetisation, 
$M_{x}$, when the multiplicity of the spin excitations included in the basis set of Eq.~(\ref{Szbasis}) is changed. The study is 
performed on a 2$\times$2$\times$2 super-cell of the VO(acac)$_{2}$ crystal, containing a total of 16 independent and 
interacting spins. From the figure it is possible to see how $\langle M_{x} \rangle$ for a given spin initialised at its maximum 
value after the $\pi/2$ pulse, decays with a Gaussian profile
\begin{equation}
M_{x}(t)=(M_{x}(0)-M_{x}(\infty))e^{-(\frac{t}{T_{2}})^{2}}+M_{x}(\infty)\:.
\label{gauss}
\end{equation}
The calculated time constant is in the nanosecond range. The short-time spin dynamics converges very rapidly with respect 
to the inclusion of multiple spin-flips in the basis set. In fact, already for the minimal basis set, including only double spin-flips, 
the decay part of the dynamics is well converged and only long-time features need to be perfected. This is an important aspect 
as it allows us to use the minimal basis-set to study the dephasing process, being the short-time dynamics the important one.  

\begin{figure}[h!]
\includegraphics[scale=1]{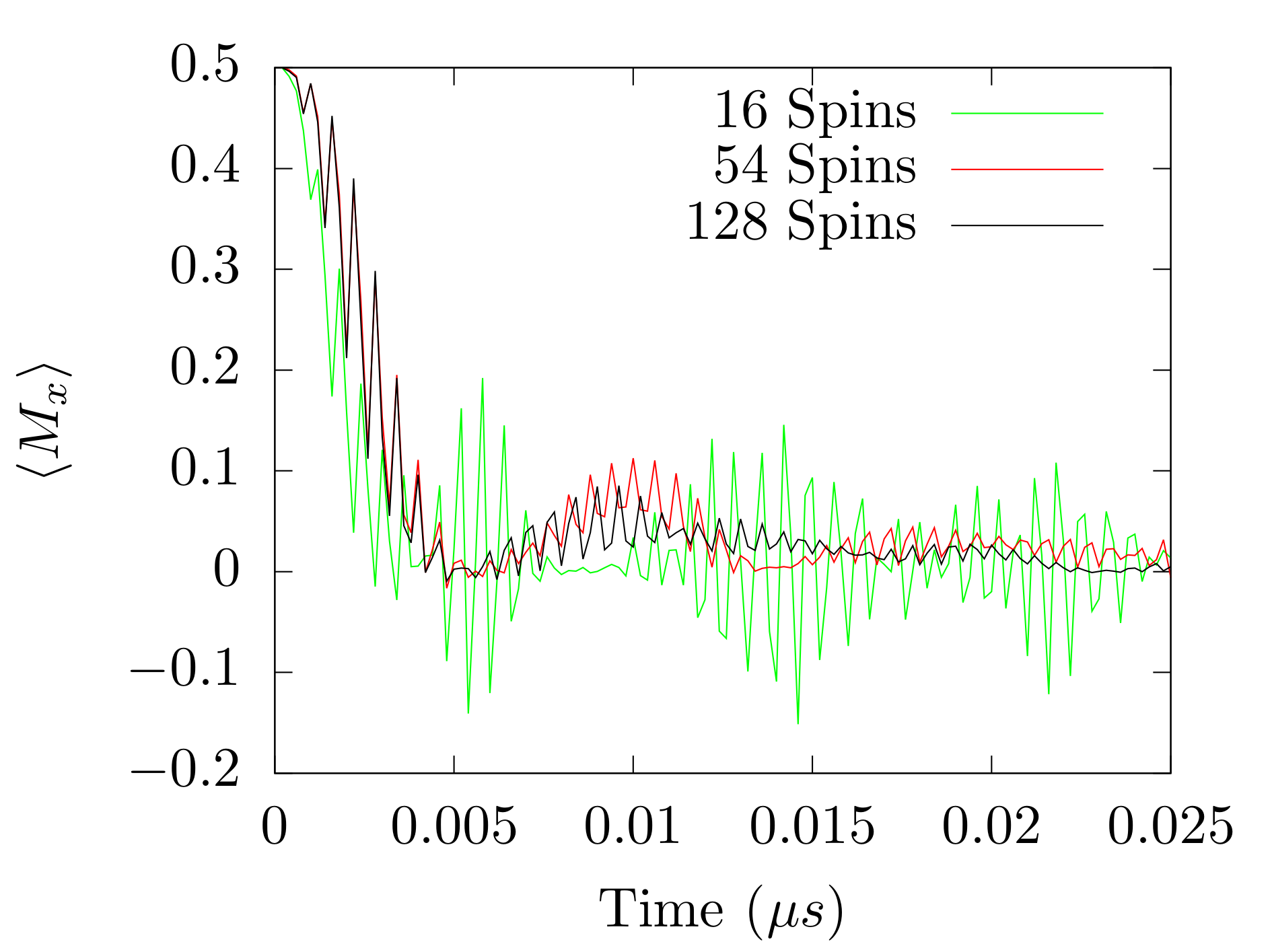}
\caption{\textbf{Spin-echo dependence on the number of spins.} The echo profiles obtained from the simulation of different sizes 
of the VO(acac)$_{2}$ supercell. These are calculated with a two spin-flips basis sets. The green curve corresponds to a 
2$\times$2$\times$2 supercell, the red one to a 3$\times$3$\times$3 and the black one to a 4$\times$4$\times$4.}
\label{echonsp}
\end{figure}

Strong of these results we use the two spin-flips basis-set to study the dependence of the $\langle M_{x} \rangle$ dynamics 
on the total number of spins included in the simulation. This is reported in Fig.~\ref{echonsp}. The spin dynamics features 
converge fast with respect to the number of interacting spins, and virtually no difference is observed on the short-time profile 
for 3$\times$3$\times$3 and 4$\times$4$\times$4 VO(acac)$_{2}$ super-cells, containing 54 and 128 spins, respectively. It is 
also interesting to note how the long-time fluctuations get damped with the inclusion of more spins in the simulation. This happens 
as a result of an increased size of the Hilbert space, which makes the chance of the system to visit states with a large magnetisation 
less probable, in analogy to the classical recurrent time in the phase space. 

\begin{figure}[h!]
\includegraphics[scale=1]{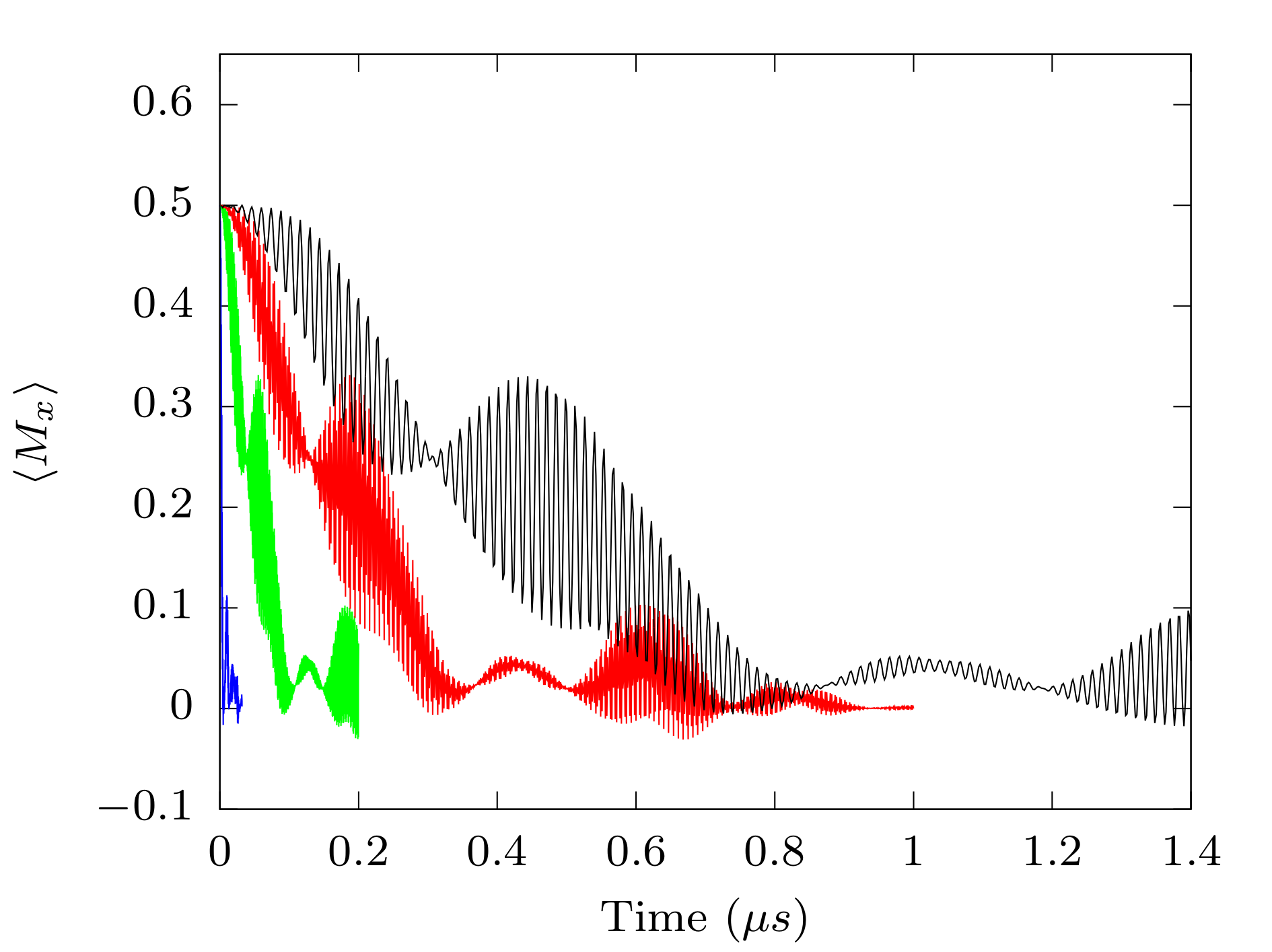}
\caption{\textbf{Spin echo dependence on the magnetic dilution.} The echo profile obtained for 27 spins arranged according 
to the VO(acac)$_{2}$ crystal symmetry is reported for different values of the dilution: 50\% (red curve), 6.25\% (green curve), 
1.85\% (red curve) and 0.78\% (black curve).}
\label{echodil}
\end{figure}

Being the limits of our method understood, we are now ready to study the effects of the magnetic dilution on the transverse 
magnetisation decaying profile. A simple model for addressing this problem consists in designing a lattice of spins with the 
same symmetry of the crystal unit-cell and rescale the size of the lattice vectors in order to modulate the spin-spin distance 
as in a dilution experiment. By applying this methodology to 27 VO(acac)$_{2}$ spins, we scan magnetic dilution values ranging from 
50\% to 0.78\% and the results are reported in Fig.~(\ref{echodil}). The dilution percentage is calculated as the number of spins with respect the total number of Vanadium sites availble in the volume considered.
The echo signal presents the same features regardless of the dilution, since the symmetry of the interactions is preserved. 
However, the time-scale changes, leading to larger $T_{2}$ values for diluted systems. These reach values of 0.1~$\mu s$ for 
$\sim1\%$ dilution.

\begin{figure}[h!]
\includegraphics[scale=1]{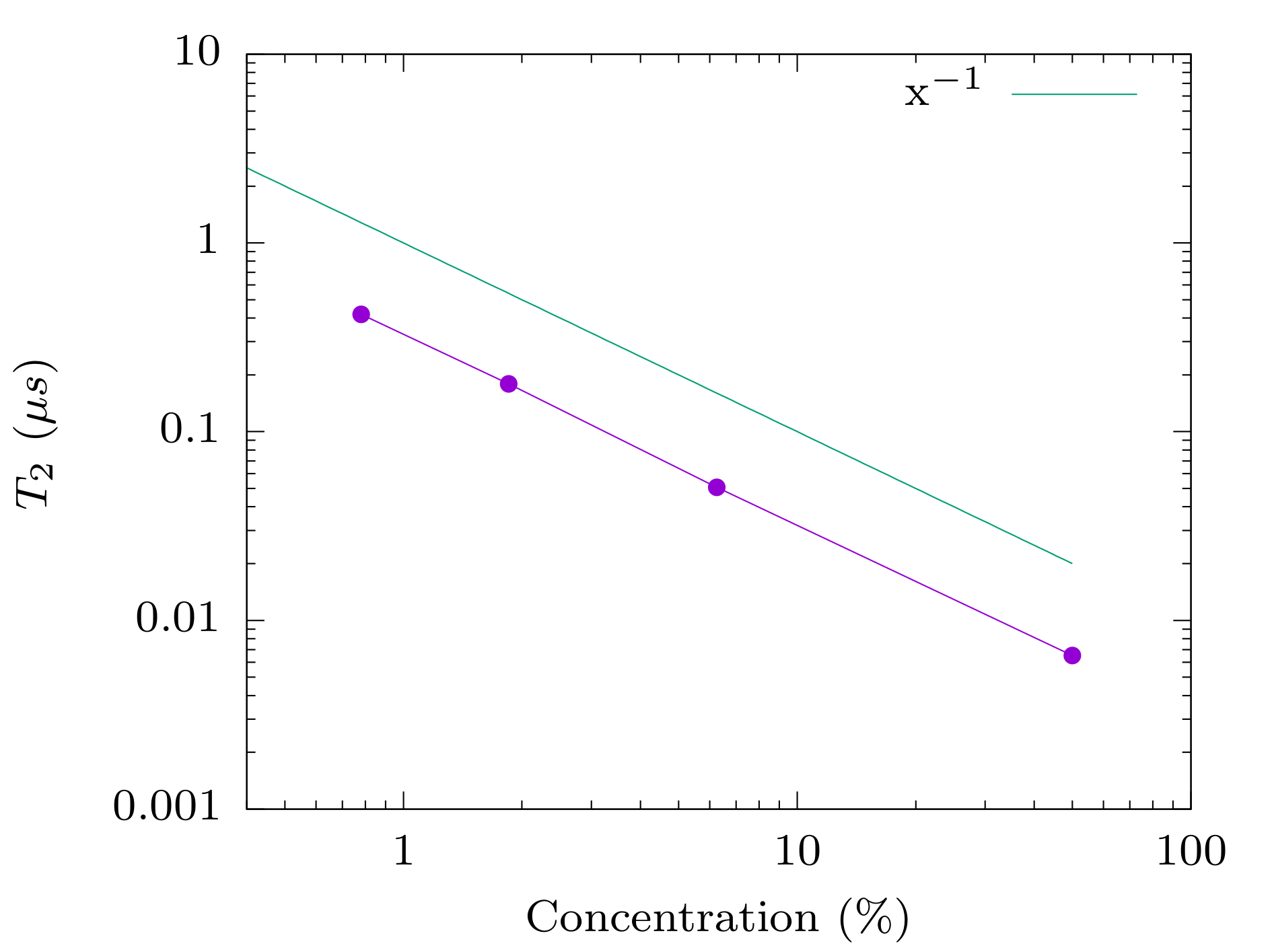}
\caption{\textbf{T$_{2}$ dependence on the magnetic dilution.} The dependence of $T_{2}$, extracted by fitting to the expression 
in Eq.~(\ref{gauss}), is shown as function of dilution percentage (purple dots and line). The green line serves as guide to explicit the x$^{-1}$ dependence.}
\label{echodil2}
\end{figure}

Fig.~\ref{echodil2} shows the $T_{2}$, extracted by fitting the echo profiles to Eq.~(\ref{gauss}), 
as a function of the dilution, revealing a linear dependence. This fact has also been previously noted in the context of nuclear 
magnetic impurities interacting with an electronic spin~\cite{Maze2008} and in diluted radicals~\cite{Brown1973}. It originates 
from the fact that, even though the dipolar interactions goes as $R^{-3}$, the average distance between spins does not scale 
linearly with the dilution percentage. 

Interestingly, even though the predicted order of magnitude of the spin dephasing is approaching the correct value for the 
typical dilutions used in experiments, the shape of the spin-echo envelope is qualitatively different. Spin-echo signals are 
usually interpreted with generalised exponentials as 
\begin{equation}
M_{x}(t)=(M_{x}(0)-M_{x}(\infty))e^{-(\frac{t}{T_{2}})^{\beta}}+M_{x}(\infty)\:.
\label{genexp}
\end{equation}
For this class of molecular complexes an experimental value $\beta<1$ is found~\cite{Atzori2016}, while here the profile is 
Gaussian and $\beta=2$.

In order to investigate this discrepancy we turn our attention to the study of another molecule in the Vanadyl family,
namely VO(dmit)$_{2}$. This compound has been extensively investigated experimentally~\cite{Atzori2016b} and the 
spin-echo decay profile is available. Furthermore, an experimental $\mathbf{g}$ tensor has been extracted and it is 
used here: $g_{xx}=g_{yy}=1.9843$ and $g_{zz}=1.9735$.
In this second example we overcome the limitations of the previous diluted study by designing several 
super-cells, where the Vanadium sites are randomly occupied by a number of spins consistent with the desired level 
of dilution. A first set of 80 supercells contains 48 spins located on the sites of a 6$\times$8$\times$6 VO(dmit)$_{2}$ 
supercell, and a second set of 80 contains 70 spins located on the sites of a 7$\times$9$\times$7 supercell. All the 
supercells correspond to a dilution of 4\%, a value comparable with experimental conditions. The spin echoes for all the 
different simulations performed on the same number of spins are then averaged. This protocol offers a more realistic 
representation of the experimental conditions, where spins experience different environments. 

\begin{figure}[h!]
\includegraphics[scale=1]{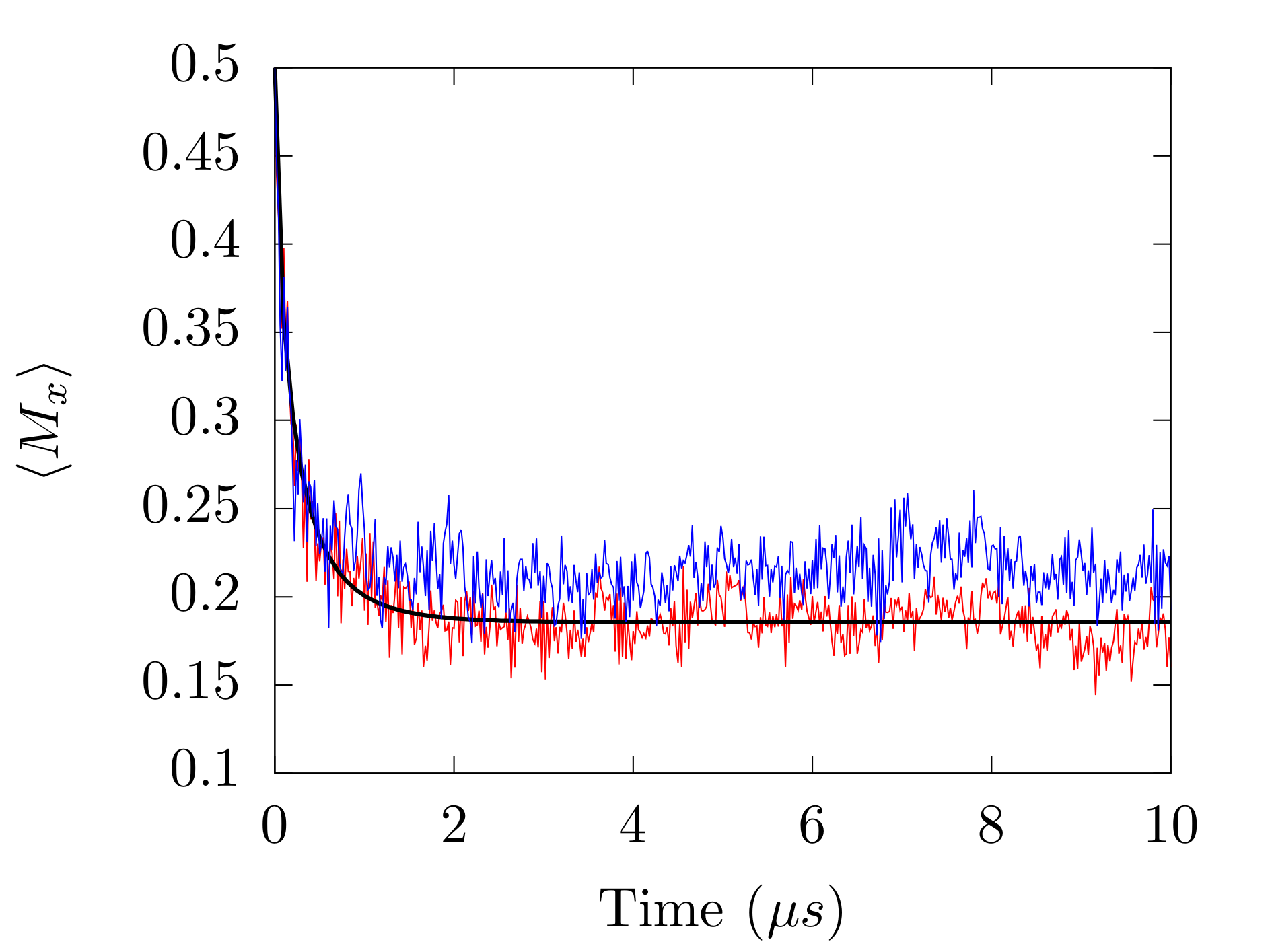}
\caption{\textbf{Spin echo dependence on structural disorder.} The time-decay profile of the $x$-component of the magnetisation 
is reported for two different sets of simulations, both corresponding to a 4\% of dilution. The average of 80 simulations performed 
with 48 spins is reported in blue, while the red curve describes the average of 80 simulations with 70 spins. The solid black line 
corresponds to the best fit of the red curve to Eq.~(\ref{genexp}).}
\label{echoave}
\end{figure}

Our results are summarised in Fig.~\ref{echoave}, where we show a non-Gaussian decaying profile of $M_{x}$ for both 
the sets of supercells, with the only noticeable difference among them being the value of $M_{x}(\infty)$. The fit to 
Eq.~(\ref{genexp}) of the curve corresponding to the simulation with 70 spins returns $\beta=0.70$ and $T_{2}=0.20~\mu s$. 
In comparison, VO(acac)$_{2}$ relaxes with $T_{2}\sim 0.10~\mu s$ for the same level of dilution, as extrapolated from Fig.\ref{echodil2}.
Notably, the stretch factor $\beta$ is now in excellent agreement with the experimental phenomenology. At the same time
we predict a $T_2$ slightly shorter than the experimental one, which is estimated to be $\sim5\mu s$. This is an interesting 
result, since usually an incomplete model overestimates the relaxation time, while our predictions surpass the experimental 
rate of decoherence. This suggests that there is an active mechanism, not cover in our simulations, slowing down the dephasing
process. However, other sources of error might may also come from the simulated pulses being instantaneous and from 
the fact that in experiments multiple spins resonate at the same frequency and would flip at once under micro-wave pulses.
Our model instead assumes that one single electronic spin is affected by the pulse.

\section{Discussion}

The study of spin dephasing in molecular magnets has been mainly driven by experimental investigation and over the years 
several relevant figures of merit for the optimisation of $T_2$ have been proposed. However, the main focus of these investigations 
concern the role of the H$^{1}$ spins in the solid-state matrix. The physics of this interaction is relatively well understood and 
the basic guideline to enhance $T_2$ consists in synthesising organic ligands and crystal's matrices without elements with 
abundant spin-active nuclear isotopes. The role of the residual electronic spins interactions is usually less investigated and 
dilution is generally thought to be sufficient to eliminate it. 

Here, however, we have shown that $T_{2}$ is only linearly dependent on the dilution and extremely low concentrations are 
needed to bring it beyond the $\mu s$ boundary. According to Fig.~\ref{echodil}, a 0.01\% dilution (or 0.3 mM) is needed to 
obtain $T_{2}>10\mu s$. This estimate also sets the minimal distance between qubits inside a device to 15~nm. For levels 
of dilution lower than this threshold, the electronic spin-spin contribution is comparable to that of H$^{1}$. Therefore, the method 
proposed here, is expected to be of help in the discrimination processes originating from different interactions. A less investigated 
aspect, due to the synthetic complexity in gauging such interaction, concerns the role of the hyperfine coupling of the nuclear spin 
borne by the same atom hosting the electronic one. The method presented in this work can also be applied to this situation, as 
well as to the central electronic spin interacting with an H$^{1}$ spin bath, and it will be argument of future applications. Another 
important ingredient that can be added to our model, is the inclusion of temperature effects and spin-phonon relaxation. As we 
approach ambient temperature, spin-phonon coupling becomes the bottle-neck for the coherence time~\cite{Bader2014,Atzori2016b}. 
In particular the understanding of the correlation between the chemical structure and the spin-lattice relaxation might lead to 
new, unforeseen synthetic designing rules~\cite{Atzori2017,Lunghi2017,Lunghi2017a} and it must be considered a research 
priority.

\section{Conclusions}

In conclusion, we have here presented a general methodology to generate a truncated Hilbert space, designed to reproduce 
the dynamics of interacting spins in external magnetic fields. The method has been tested for VO(acac)$_{2}$ and VO(dmit)$_{2}$ 
molecular qubits, showing that even simple two-spin flips and second neighbour shells are able to capture the relevant physics 
leading to spin dephasing. The study of VO(acac)$_{2}$ has also been extended to reproduce magnetic dilution experiments 
and has shown a linear dependence of $T_{2}$ with respect to the dilution. Our results therefore highlight the important role 
played by the residual electronic spin dipolar interaction in crystals of molecular qubits in setting a upper limit to $T_{2}$, 
reachable at not-extreme dilutions. Finally, our simulations also shed some light on the origin of the stretched exponential-type 
of decay usually encountered in EPR investigations of transition metal complexes, pointing to the disorder of magnetic centres 
inside the solid-state host as its origin. \\

\noindent
\textbf{Data Availability}\\
All the relevant data discussed in the present paper are available from the authors upon request. \\

\noindent
\textbf{Acknowledgments}\\
This work has been sponsored by Science Foundation Ireland (grant 14/IA/2624). 
We acknowledge the MOLSPIN COST action CA15128. Computational resources were provided by the Trinity 
Centre for High Performance Computing (TCHPC) and the Irish Centre for High-End Computing (ICHEC). \\

\noindent
\textbf{Author contributions}\\
All the authors discussed the results and contributed to the manuscript. \\

\noindent
\textbf{Competing financial interests}\\
The authors declare no competing financial interests.


\begin{thebibliography}{31}%
\makeatletter
\providecommand \@ifxundefined [1]{%
 \@ifx{#1\undefined}
}%
\providecommand \@ifnum [1]{%
 \ifnum #1\expandafter \@firstoftwo
 \else \expandafter \@secondoftwo
 \fi
}%
\providecommand \@ifx [1]{%
 \ifx #1\expandafter \@firstoftwo
 \else \expandafter \@secondoftwo
 \fi
}%
\providecommand \natexlab [1]{#1}%
\providecommand \enquote  [1]{``#1''}%
\providecommand \bibnamefont  [1]{#1}%
\providecommand \bibfnamefont [1]{#1}%
\providecommand \citenamefont [1]{#1}%
\providecommand \href@noop [0]{\@secondoftwo}%
\providecommand \href [0]{\begingroup \@sanitize@url \@href}%
\providecommand \@href[1]{\@@startlink{#1}\@@href}%
\providecommand \@@href[1]{\endgroup#1\@@endlink}%
\providecommand \@sanitize@url [0]{\catcode `\\12\catcode `\$12\catcode
  `\&12\catcode `\#12\catcode `\^12\catcode `\_12\catcode `\%12\relax}%
\providecommand \@@startlink[1]{}%
\providecommand \@@endlink[0]{}%
\providecommand \url  [0]{\begingroup\@sanitize@url \@url }%
\providecommand \@url [1]{\endgroup\@href {#1}{\urlprefix }}%
\providecommand \urlprefix  [0]{URL }%
\providecommand \Eprint [0]{\href }%
\providecommand \doibase [0]{http://dx.doi.org/}%
\providecommand \selectlanguage [0]{\@gobble}%
\providecommand \bibinfo  [0]{\@secondoftwo}%
\providecommand \bibfield  [0]{\@secondoftwo}%
\providecommand \translation [1]{[#1]}%
\providecommand \BibitemOpen [0]{}%
\providecommand \bibitemStop [0]{}%
\providecommand \bibitemNoStop [0]{.\EOS\space}%
\providecommand \EOS [0]{\spacefactor3000\relax}%
\providecommand \BibitemShut  [1]{\csname bibitem#1\endcsname}%
\let\auto@bib@innerbib\@empty
\bibitem [{\citenamefont {Nakamura}\ \emph {et~al.}(1999)\citenamefont
  {Nakamura}, \citenamefont {Pashkin},\ and\ \citenamefont
  {Tsai}}]{Nakamura1999}%
  \BibitemOpen
  \bibfield  {author} {\bibinfo {author} {\bibfnamefont {Y}~\bibnamefont
  {Nakamura}}, \bibinfo {author} {\bibfnamefont {Y~A}\ \bibnamefont {Pashkin}},
  \ and\ \bibinfo {author} {\bibfnamefont {J~S}\ \bibnamefont {Tsai}},\
  }\bibfield  {title} {\enquote {\bibinfo {title} {{Coherent Control of
  Macroscopic quantum states in a single cooper pair box}},}\ }\href {\doibase
  10.1038/19718} {\bibfield  {journal} {\bibinfo  {journal} {Nature}\ }\textbf
  {\bibinfo {volume} {398}},\ \bibinfo {pages} {786--788} (\bibinfo {year}
  {1999})}\BibitemShut {NoStop}%
\bibitem [{\citenamefont {Godfrin}\ \emph {et~al.}(2017)\citenamefont
  {Godfrin}, \citenamefont {Ferhat}, \citenamefont {Ballou}, \citenamefont
  {Klyatskaya}, \citenamefont {Ruben}, \citenamefont {Wernsdorfer},\ and\
  \citenamefont {Balestro}}]{Godfrin2017}%
  \BibitemOpen
  \bibfield  {author} {\bibinfo {author} {\bibfnamefont {C.}~\bibnamefont
  {Godfrin}}, \bibinfo {author} {\bibfnamefont {A.}~\bibnamefont {Ferhat}},
  \bibinfo {author} {\bibfnamefont {R.}~\bibnamefont {Ballou}}, \bibinfo
  {author} {\bibfnamefont {S.}~\bibnamefont {Klyatskaya}}, \bibinfo {author}
  {\bibfnamefont {M.}~\bibnamefont {Ruben}}, \bibinfo {author} {\bibfnamefont
  {W.}~\bibnamefont {Wernsdorfer}}, \ and\ \bibinfo {author} {\bibfnamefont
  {F.}~\bibnamefont {Balestro}},\ }\bibfield  {title} {\enquote {\bibinfo
  {title} {{Operating Quantum States in Single Magnetic Molecules:
  Implementation of Grover's Quantum Algorithm}},}\ }\href@noop {} {\bibfield
  {journal} {\bibinfo  {journal} {Phys. Rev. Lett.}\ }\textbf {\bibinfo
  {volume} {119}},\ \bibinfo {pages} {187702} (\bibinfo {year}
  {2017})}\BibitemShut {NoStop}%
\bibitem [{\citenamefont {DiVincenzo}(2000)}]{DiVincenzo2000}%
  \BibitemOpen
  \bibfield  {author} {\bibinfo {author} {\bibfnamefont {David~P.}\
  \bibnamefont {DiVincenzo}},\ }\bibfield  {title} {\enquote {\bibinfo {title}
  {{The Physical Implementation of Quantum Computing}},}\ }\href {\doibase
  10.1029/91WR02224} {\bibfield  {journal} {\bibinfo  {journal} {Fortschr.
  Phys.}\ }\textbf {\bibinfo {volume} {48}},\ \bibinfo {pages} {771--783}
  (\bibinfo {year} {2000})}\BibitemShut {NoStop}%
\bibitem [{\citenamefont {Atzori}\ \emph
  {et~al.}(2016{\natexlab{a}})\citenamefont {Atzori}, \citenamefont {Morra},
  \citenamefont {Tesi}, \citenamefont {Albino}, \citenamefont {Chiesa},
  \citenamefont {Sorace},\ and\ \citenamefont {Sessoli}}]{Atzori2016b}%
  \BibitemOpen
  \bibfield  {author} {\bibinfo {author} {\bibfnamefont {Matteo}\ \bibnamefont
  {Atzori}}, \bibinfo {author} {\bibfnamefont {Elena}\ \bibnamefont {Morra}},
  \bibinfo {author} {\bibfnamefont {Lorenzo}\ \bibnamefont {Tesi}}, \bibinfo
  {author} {\bibfnamefont {Andrea}\ \bibnamefont {Albino}}, \bibinfo {author}
  {\bibfnamefont {Mario}\ \bibnamefont {Chiesa}}, \bibinfo {author}
  {\bibfnamefont {Lorenzo}\ \bibnamefont {Sorace}}, \ and\ \bibinfo {author}
  {\bibfnamefont {Roberta}\ \bibnamefont {Sessoli}},\ }\bibfield  {title}
  {\enquote {\bibinfo {title} {{Quantum Coherence Times Enhancement in
  Vanadium(IV)-based Potential Molecular Qubits: The Key Role of the Vanadyl
  Moiety}},}\ }\href {\doibase 10.1021/jacs.6b05574} {\bibfield  {journal}
  {\bibinfo  {journal} {J. Am. Chem. Soc.}\ }\textbf {\bibinfo {volume}
  {138}},\ \bibinfo {pages} {11234--11244} (\bibinfo {year}
  {2016}{\natexlab{a}})}\BibitemShut {NoStop}%
\bibitem [{\citenamefont {Zadrozny}\ \emph {et~al.}(2015)\citenamefont
  {Zadrozny}, \citenamefont {Niklas}, \citenamefont {Poluektov},\ and\
  \citenamefont {Freedman}}]{Zadrozny2015}%
  \BibitemOpen
  \bibfield  {author} {\bibinfo {author} {\bibfnamefont {Joseph~M.}\
  \bibnamefont {Zadrozny}}, \bibinfo {author} {\bibfnamefont {Jens}\
  \bibnamefont {Niklas}}, \bibinfo {author} {\bibfnamefont {Oleg~G.}\
  \bibnamefont {Poluektov}}, \ and\ \bibinfo {author} {\bibfnamefont
  {Danna~E.}\ \bibnamefont {Freedman}},\ }\bibfield  {title} {\enquote
  {\bibinfo {title} {{Millisecond Coherence Time in a Tunable Molecular
  Electronic Spin Qubit}},}\ }\href {\doibase 10.1021/acscentsci.5b00338}
  {\bibfield  {journal} {\bibinfo  {journal} {ACS Cent. Sci.}\ }\textbf
  {\bibinfo {volume} {1}},\ \bibinfo {pages} {488--492} (\bibinfo {year}
  {2015})}\BibitemShut {NoStop}%
\bibitem [{\citenamefont {Leuenberger}\ and\ \citenamefont
  {Loss}(2001)}]{Leuenberger2001}%
  \BibitemOpen
  \bibfield  {author} {\bibinfo {author} {\bibfnamefont {Michael~N.}\
  \bibnamefont {Leuenberger}}\ and\ \bibinfo {author} {\bibfnamefont {Daniel}\
  \bibnamefont {Loss}},\ }\bibfield  {title} {\enquote {\bibinfo {title}
  {{Quantum computing in molecular magnets.}}}\ }\href {\doibase
  10.1038/35071024} {\bibfield  {journal} {\bibinfo  {journal} {Nature}\
  }\textbf {\bibinfo {volume} {410}},\ \bibinfo {pages} {789--93} (\bibinfo
  {year} {2001})}\BibitemShut {NoStop}%
\bibitem [{\citenamefont {Ardavan}\ \emph {et~al.}(2007)\citenamefont
  {Ardavan}, \citenamefont {Rival}, \citenamefont {Morton}, \citenamefont
  {Blundell}, \citenamefont {Tyryshkin}, \citenamefont {Timco},\ and\
  \citenamefont {Winpenny}}]{Ardavan2007}%
  \BibitemOpen
  \bibfield  {author} {\bibinfo {author} {\bibfnamefont {Arzhang}\ \bibnamefont
  {Ardavan}}, \bibinfo {author} {\bibfnamefont {Olivier}\ \bibnamefont
  {Rival}}, \bibinfo {author} {\bibfnamefont {John J.~L.}\ \bibnamefont
  {Morton}}, \bibinfo {author} {\bibfnamefont {Stephen~J.}\ \bibnamefont
  {Blundell}}, \bibinfo {author} {\bibfnamefont {Alexei~M.}\ \bibnamefont
  {Tyryshkin}}, \bibinfo {author} {\bibfnamefont {Grigore~a.}\ \bibnamefont
  {Timco}}, \ and\ \bibinfo {author} {\bibfnamefont {Richard E.~P.}\
  \bibnamefont {Winpenny}},\ }\bibfield  {title} {\enquote {\bibinfo {title}
  {{Will Spin-Relaxation Times in Molecular Magnets Permit Quantum Information
  Processing?}}}\ }\href {\doibase 10.1103/PhysRevLett.98.057201} {\bibfield
  {journal} {\bibinfo  {journal} {Phys. Rev. Lett.}\ }\textbf {\bibinfo
  {volume} {98}},\ \bibinfo {pages} {057201} (\bibinfo {year}
  {2007})}\BibitemShut {NoStop}%
\bibitem [{\citenamefont {Winpenny}(2008)}]{Winpenny2008}%
  \BibitemOpen
  \bibfield  {author} {\bibinfo {author} {\bibfnamefont {Richard E~P}\
  \bibnamefont {Winpenny}},\ }\bibfield  {title} {\enquote {\bibinfo {title}
  {{Quantum Information Processing Using Molecular Nanomagnets As Qubits}},}\
  }\href {\doibase 10.1002/anie.200802742} {\bibfield  {journal} {\bibinfo
  {journal} {Angew. Chem. Int. Ed.}\ }\textbf {\bibinfo {volume} {47}},\
  \bibinfo {pages} {7992--7994} (\bibinfo {year} {2008})}\BibitemShut {NoStop}%
\bibitem [{\citenamefont {Lehmann}\ \emph {et~al.}(2009)\citenamefont
  {Lehmann}, \citenamefont {Gaita-Ari{\~{n}}o}, \citenamefont {Coronado},\ and\
  \citenamefont {Loss}}]{Lehmann2009}%
  \BibitemOpen
  \bibfield  {author} {\bibinfo {author} {\bibfnamefont {J}~\bibnamefont
  {Lehmann}}, \bibinfo {author} {\bibfnamefont {A}~\bibnamefont
  {Gaita-Ari{\~{n}}o}}, \bibinfo {author} {\bibfnamefont {E}~\bibnamefont
  {Coronado}}, \ and\ \bibinfo {author} {\bibfnamefont {D}~\bibnamefont
  {Loss}},\ }\bibfield  {title} {\enquote {\bibinfo {title} {{Quantum computing
  with molecular spin systems †}},}\ }\href {\doibase 10.1039/b810634g}
  {\bibfield  {journal} {\bibinfo  {journal} {J. Mater. Chem.}\ }\textbf
  {\bibinfo {volume} {19}},\ \bibinfo {pages} {1672--1677} (\bibinfo {year}
  {2009})}\BibitemShut {NoStop}%
\bibitem [{\citenamefont {Wedge}\ \emph {et~al.}(2012)\citenamefont {Wedge},
  \citenamefont {Timco}, \citenamefont {Spielberg}, \citenamefont {George},
  \citenamefont {Tuna}, \citenamefont {Rigby}, \citenamefont {McInnes},
  \citenamefont {Winpenny}, \citenamefont {Blundell},\ and\ \citenamefont
  {Ardavan}}]{Wedge2012}%
  \BibitemOpen
  \bibfield  {author} {\bibinfo {author} {\bibfnamefont {C.~J.}\ \bibnamefont
  {Wedge}}, \bibinfo {author} {\bibfnamefont {G.~A.}\ \bibnamefont {Timco}},
  \bibinfo {author} {\bibfnamefont {E.~T.}\ \bibnamefont {Spielberg}}, \bibinfo
  {author} {\bibfnamefont {R.~E.}\ \bibnamefont {George}}, \bibinfo {author}
  {\bibfnamefont {F.}~\bibnamefont {Tuna}}, \bibinfo {author} {\bibfnamefont
  {S.}~\bibnamefont {Rigby}}, \bibinfo {author} {\bibfnamefont {E.~J.L.}\
  \bibnamefont {McInnes}}, \bibinfo {author} {\bibfnamefont {R.~E.P.}\
  \bibnamefont {Winpenny}}, \bibinfo {author} {\bibfnamefont {S.~J.}\
  \bibnamefont {Blundell}}, \ and\ \bibinfo {author} {\bibfnamefont
  {A.}~\bibnamefont {Ardavan}},\ }\bibfield  {title} {\enquote {\bibinfo
  {title} {{Chemical engineering of molecular qubits}},}\ }\href {\doibase
  10.1103/PhysRevLett.108.107204} {\bibfield  {journal} {\bibinfo  {journal}
  {Phys. Rev. Lett.}\ }\textbf {\bibinfo {volume} {108}},\ \bibinfo {pages}
  {107204} (\bibinfo {year} {2012})}\BibitemShut {NoStop}%
\bibitem [{\citenamefont {Ferrando-Soria}\ \emph {et~al.}(2016)\citenamefont
  {Ferrando-Soria}, \citenamefont {Pineda}, \citenamefont {Chiesa},
  \citenamefont {Fernandez}, \citenamefont {Magee}, \citenamefont {Carretta},
  \citenamefont {Santini}, \citenamefont {Vitorica-Yrezabal}, \citenamefont
  {Tuna}, \citenamefont {Timco}, \citenamefont {Mcinnes}, \citenamefont
  {Carretta}, \citenamefont {Santini},\ and\ \citenamefont
  {Winpenny}}]{Ferrando-Soria2016}%
  \BibitemOpen
  \bibfield  {author} {\bibinfo {author} {\bibfnamefont {J}~\bibnamefont
  {Ferrando-Soria}}, \bibinfo {author} {\bibfnamefont {E~M}\ \bibnamefont
  {Pineda}}, \bibinfo {author} {\bibfnamefont {A}~\bibnamefont {Chiesa}},
  \bibinfo {author} {\bibfnamefont {A}~\bibnamefont {Fernandez}}, \bibinfo
  {author} {\bibfnamefont {S~A}\ \bibnamefont {Magee}}, \bibinfo {author}
  {\bibfnamefont {S}~\bibnamefont {Carretta}}, \bibinfo {author} {\bibfnamefont
  {P}~\bibnamefont {Santini}}, \bibinfo {author} {\bibfnamefont
  {J}~\bibnamefont {Vitorica-Yrezabal}}, \bibinfo {author} {\bibfnamefont
  {F}~\bibnamefont {Tuna}}, \bibinfo {author} {\bibfnamefont {G~A}\
  \bibnamefont {Timco}}, \bibinfo {author} {\bibfnamefont {E~J~L}\ \bibnamefont
  {Mcinnes}}, \bibinfo {author} {\bibfnamefont {S}~\bibnamefont {Carretta}},
  \bibinfo {author} {\bibfnamefont {P}~\bibnamefont {Santini}}, \ and\ \bibinfo
  {author} {\bibfnamefont {R~E~P}\ \bibnamefont {Winpenny}},\ }\bibfield
  {title} {\enquote {\bibinfo {title} {{A modular design of molecular qubits to
  implement universal quantum gates}},}\ }\href {\doibase 10.1038/ncomms11377}
  {\bibfield  {journal} {\bibinfo  {journal} {Nat. Comm.}\ }\textbf {\bibinfo
  {volume} {7}},\ \bibinfo {pages} {11377} (\bibinfo {year}
  {2016})}\BibitemShut {NoStop}%
\bibitem [{\citenamefont {Shiddiq}\ \emph {et~al.}(2016)\citenamefont
  {Shiddiq}, \citenamefont {Komijani}, \citenamefont {Duan}, \citenamefont
  {Gaita-Ari{\~{n}}o}, \citenamefont {Coronado},\ and\ \citenamefont
  {Hill}}]{Shiddiq2016}%
  \BibitemOpen
  \bibfield  {author} {\bibinfo {author} {\bibfnamefont {Muhandis}\
  \bibnamefont {Shiddiq}}, \bibinfo {author} {\bibfnamefont {Dorsa}\
  \bibnamefont {Komijani}}, \bibinfo {author} {\bibfnamefont {Yan}\
  \bibnamefont {Duan}}, \bibinfo {author} {\bibfnamefont {Alejandro}\
  \bibnamefont {Gaita-Ari{\~{n}}o}}, \bibinfo {author} {\bibfnamefont
  {Eugenio}\ \bibnamefont {Coronado}}, \ and\ \bibinfo {author} {\bibfnamefont
  {Stephen}\ \bibnamefont {Hill}},\ }\bibfield  {title} {\enquote {\bibinfo
  {title} {{Enhancing coherence in molecular spin qubits via atomic clock
  transitions}},}\ }\href {\doibase 10.1038/nature16984} {\bibfield  {journal}
  {\bibinfo  {journal} {Nature}\ }\textbf {\bibinfo {volume} {531}},\ \bibinfo
  {pages} {348--351} (\bibinfo {year} {2016})}\BibitemShut {NoStop}%
\bibitem [{\citenamefont {Gaita-Ari{\~{n}}o}\ \emph {et~al.}(2019)\citenamefont
  {Gaita-Ari{\~{n}}o}, \citenamefont {Luis}, \citenamefont {Hill},\ and\
  \citenamefont {Coronado}}]{Gaita-Arino2019}%
  \BibitemOpen
  \bibfield  {author} {\bibinfo {author} {\bibfnamefont {A}~\bibnamefont
  {Gaita-Ari{\~{n}}o}}, \bibinfo {author} {\bibfnamefont {F}~\bibnamefont
  {Luis}}, \bibinfo {author} {\bibfnamefont {S}~\bibnamefont {Hill}}, \ and\
  \bibinfo {author} {\bibfnamefont {E}~\bibnamefont {Coronado}},\ }\bibfield
  {title} {\enquote {\bibinfo {title} {{Molecular spins for quantum
  computation}},}\ }\href {\doibase 10.1038/s41557-019-0232-y} {\bibfield
  {journal} {\bibinfo  {journal} {Nat. Chem.}\ }\textbf {\bibinfo {volume}
  {11}},\ \bibinfo {pages} {301--309} (\bibinfo {year} {2019})}\BibitemShut
  {NoStop}%
\bibitem [{\citenamefont {Yao}\ \emph {et~al.}(2006)\citenamefont {Yao},
  \citenamefont {Liu},\ and\ \citenamefont {Sham}}]{Yao2006}%
  \BibitemOpen
  \bibfield  {author} {\bibinfo {author} {\bibfnamefont {Wang}\ \bibnamefont
  {Yao}}, \bibinfo {author} {\bibfnamefont {Ren~Bao}\ \bibnamefont {Liu}}, \
  and\ \bibinfo {author} {\bibfnamefont {L.~J.}\ \bibnamefont {Sham}},\
  }\bibfield  {title} {\enquote {\bibinfo {title} {{Theory of electron spin
  decoherence by interacting nuclear spins in a quantum dot}},}\ }\href
  {\doibase 10.1103/PhysRevB.74.195301} {\bibfield  {journal} {\bibinfo
  {journal} {Phys. Rev. B}\ }\textbf {\bibinfo
  {volume} {74}},\ \bibinfo {pages} {195301} (\bibinfo {year}
  {2006})}\BibitemShut {NoStop}%
\bibitem [{\citenamefont {Witzel}\ and\ \citenamefont {{Das
  Sarma}}(2006)}]{Witzel2006}%
  \BibitemOpen
  \bibfield  {author} {\bibinfo {author} {\bibfnamefont {W.~M.}\ \bibnamefont
  {Witzel}}\ and\ \bibinfo {author} {\bibfnamefont {S.}~\bibnamefont {{Das
  Sarma}}},\ }\bibfield  {title} {\enquote {\bibinfo {title} {{Quantum theory
  for electron spin decoherence induced by nuclear spin dynamics in
  semiconductor quantum computer architectures: Spectral diffusion of localized
  electron spins in the nuclear solid-state environment}},}\ }\href {\doibase
  10.1103/PhysRevB.74.035322} {\bibfield  {journal} {\bibinfo  {journal} {Phys.
  Rev. B}\ }\textbf {\bibinfo {volume} {74}},\
  \bibinfo {pages} {035322} (\bibinfo {year} {2006})}\BibitemShut {NoStop}%
\bibitem [{\citenamefont {Yang}\ and\ \citenamefont {Liu}(2008)}]{Yang2008}%
  \BibitemOpen
  \bibfield  {author} {\bibinfo {author} {\bibfnamefont {Wen}\ \bibnamefont
  {Yang}}\ and\ \bibinfo {author} {\bibfnamefont {Ren~Bao}\ \bibnamefont
  {Liu}},\ }\bibfield  {title} {\enquote {\bibinfo {title} {{Quantum many-body
  theory for qubit decoherence in a finite-size spin bath}},}\ }\href {\doibase
  10.1063/1.3037140} {\bibfield  {journal} {\bibinfo  {journal} {Phys. Rev. B}\
  }\textbf {\bibinfo {volume} {78}},\ \bibinfo {pages} {085315} (\bibinfo
  {year} {2008})}\BibitemShut {NoStop}%
\bibitem [{\citenamefont {Maze}\ \emph {et~al.}(2008)\citenamefont {Maze},
  \citenamefont {Taylor},\ and\ \citenamefont {Lukin}}]{Maze2008}%
  \BibitemOpen
  \bibfield  {author} {\bibinfo {author} {\bibfnamefont {J.~R.}\ \bibnamefont
  {Maze}}, \bibinfo {author} {\bibfnamefont {J.~M.}\ \bibnamefont {Taylor}}, \
  and\ \bibinfo {author} {\bibfnamefont {M.~D.}\ \bibnamefont {Lukin}},\
  }\bibfield  {title} {\enquote {\bibinfo {title} {{Electron spin decoherence
  of single nitrogen-vacancy defects in diamond}},}\ }\href {\doibase
  10.1103/PhysRevB.78.094303} {\bibfield  {journal} {\bibinfo  {journal} {Phys.
  Rev. B}\ }\textbf {\bibinfo {volume} {78}},\
  \bibinfo {pages} {094303} (\bibinfo {year} {2008})}\BibitemShut {NoStop}%
\bibitem [{\citenamefont {Seo}\ \emph {et~al.}(2016)\citenamefont {Seo},
  \citenamefont {Falk}, \citenamefont {Klimov}, \citenamefont {Miao},
  \citenamefont {Galli},\ and\ \citenamefont {Awschalom}}]{Seo2016}%
  \BibitemOpen
  \bibfield  {author} {\bibinfo {author} {\bibfnamefont {Hosung}\ \bibnamefont
  {Seo}}, \bibinfo {author} {\bibfnamefont {Abram~L.}\ \bibnamefont {Falk}},
  \bibinfo {author} {\bibfnamefont {Paul~V.}\ \bibnamefont {Klimov}}, \bibinfo
  {author} {\bibfnamefont {Kevin~C.}\ \bibnamefont {Miao}}, \bibinfo {author}
  {\bibfnamefont {Giulia}\ \bibnamefont {Galli}}, \ and\ \bibinfo {author}
  {\bibfnamefont {David~D.}\ \bibnamefont {Awschalom}},\ }\bibfield  {title}
  {\enquote {\bibinfo {title} {{Quantum decoherence dynamics of divacancy spins
  in silicon carbide}},}\ }\href {\doibase 10.1038/ncomms12935} {\bibfield
  {journal} {\bibinfo  {journal} {Nat. Commun.}\ }\textbf {\bibinfo {volume}
  {7}},\ \bibinfo {pages} {12935} (\bibinfo {year} {2016})}\BibitemShut
  {NoStop}%
\bibitem [{\citenamefont {Bader}\ \emph {et~al.}(2016)\citenamefont {Bader},
  \citenamefont {Winkler},\ and\ \citenamefont {{Van Slageren}}}]{Bader2016a}%
  \BibitemOpen
  \bibfield  {author} {\bibinfo {author} {\bibfnamefont {K.}~\bibnamefont
  {Bader}}, \bibinfo {author} {\bibfnamefont {M.}~\bibnamefont {Winkler}}, \
  and\ \bibinfo {author} {\bibfnamefont {J.}~\bibnamefont {{Van Slageren}}},\
  }\bibfield  {title} {\enquote {\bibinfo {title} {{Tuning of molecular qubits:
  Very long coherence and spin-lattice relaxation times}},}\ }\href {\doibase
  10.1039/c6cc00300a} {\bibfield  {journal} {\bibinfo  {journal} {Chem.
  Commun.}\ }\textbf {\bibinfo {volume} {52}},\ \bibinfo {pages} {3623--3626}
  (\bibinfo {year} {2016})}\BibitemShut {NoStop}%
\bibitem [{\citenamefont {Roger}\ and\ \citenamefont
  {Hetherington}(1990)}]{Roger1990}%
  \BibitemOpen
  \bibfield  {author} {\bibinfo {author} {\bibfnamefont {M.}~\bibnamefont
  {Roger}}\ and\ \bibinfo {author} {\bibfnamefont {J.~H.}\ \bibnamefont
  {Hetherington}},\ }\bibfield  {title} {\enquote {\bibinfo {title}
  {{Coupled-cluster approximation for spinlattices: Application to solid
  He}},}\ }\href@noop {} {\bibfield  {journal} {\bibinfo  {journal} {Phys. Rev.
  B}\ }\textbf {\bibinfo {volume} {41}},\ \bibinfo {pages} {200--219} (\bibinfo
  {year} {1990})}\BibitemShut {NoStop}%
\bibitem [{\citenamefont {Daley}\ \emph {et~al.}(2004)\citenamefont {Daley},
  \citenamefont {Kollath}, \citenamefont {Schollwoeck},\ and\ \citenamefont
  {Vidal}}]{Daley2004}%
  \BibitemOpen
  \bibfield  {author} {\bibinfo {author} {\bibfnamefont {A.~J.}\ \bibnamefont
  {Daley}}, \bibinfo {author} {\bibfnamefont {C.}~\bibnamefont {Kollath}},
  \bibinfo {author} {\bibfnamefont {U.}~\bibnamefont {Schollwoeck}}, \ and\
  \bibinfo {author} {\bibfnamefont {G.}~\bibnamefont {Vidal}},\ }\bibfield
  {title} {\enquote {\bibinfo {title} {{Time-dependent density-matrix
  renormalization-group using adaptive effective Hilbert spaces}},}\ }\href
  {\doibase 10.1088/1742-5468/2004/04/P04005} {\bibfield  {journal} {\bibinfo
  {journal} {J. Stat. Mech. Theory Exp.}\ }\textbf {\bibinfo {volume}
  {P04005}},\ \bibinfo {pages} {1--28} (\bibinfo {year} {2004})}\BibitemShut
  {NoStop}%
\bibitem [{\citenamefont {Kuprov}\ \emph {et~al.}(2007)\citenamefont {Kuprov},
  \citenamefont {Wagner-Rundell},\ and\ \citenamefont {Hore}}]{Kuprov2007}%
  \BibitemOpen
  \bibfield  {author} {\bibinfo {author} {\bibfnamefont {Ilya}\ \bibnamefont
  {Kuprov}}, \bibinfo {author} {\bibfnamefont {Nicola}\ \bibnamefont
  {Wagner-Rundell}}, \ and\ \bibinfo {author} {\bibfnamefont {P.~J.}\
  \bibnamefont {Hore}},\ }\bibfield  {title} {\enquote {\bibinfo {title}
  {{Polynomially scaling spin dynamics simulation algorithm based on adaptive
  state-space restriction}},}\ }\href {\doibase 10.1016/j.jmr.2007.09.014}
  {\bibfield  {journal} {\bibinfo  {journal} {J. Magn. Reson.}\ }\textbf
  {\bibinfo {volume} {189}},\ \bibinfo {pages} {241--250} (\bibinfo {year}
  {2007})}\BibitemShut {NoStop}%
\bibitem [{\citenamefont {Choi}\ \emph {et~al.}(1996)\citenamefont {Choi},
  \citenamefont {Demmel}, \citenamefont {Dhillon}, \citenamefont {Dongarra},
  \citenamefont {Ostrouchov}, \citenamefont {Petitet}, \citenamefont {Stanley},
  \citenamefont {Walker},\ and\ \citenamefont {Whaley}}]{Choi1996}%
  \BibitemOpen
  \bibfield  {author} {\bibinfo {author} {\bibfnamefont {J.}~\bibnamefont
  {Choi}}, \bibinfo {author} {\bibfnamefont {J.}~\bibnamefont {Demmel}},
  \bibinfo {author} {\bibfnamefont {I.}~\bibnamefont {Dhillon}}, \bibinfo
  {author} {\bibfnamefont {J.}~\bibnamefont {Dongarra}}, \bibinfo {author}
  {\bibfnamefont {S.}~\bibnamefont {Ostrouchov}}, \bibinfo {author}
  {\bibfnamefont {A.}~\bibnamefont {Petitet}}, \bibinfo {author} {\bibfnamefont
  {K.}~\bibnamefont {Stanley}}, \bibinfo {author} {\bibfnamefont
  {D.}~\bibnamefont {Walker}}, \ and\ \bibinfo {author} {\bibfnamefont {R.~C.}\
  \bibnamefont {Whaley}},\ }\bibfield  {title} {\enquote {\bibinfo {title}
  {{ScaLAPACK: A portable linear algebra library for distributed memory
  computers - Design issues and performance}},}\ }\href {\doibase
  10.1016/0010-4655(96)00017-3} {\bibfield  {journal} {\bibinfo  {journal}
  {Comput. Phys. Commun.}\ }\textbf {\bibinfo {volume} {97}},\ \bibinfo {pages}
  {1--15} (\bibinfo {year} {1996})}\BibitemShut {NoStop}%
\bibitem [{\citenamefont {Neese}(2012)}]{Neese2012}%
  \BibitemOpen
  \bibfield  {author} {\bibinfo {author} {\bibfnamefont {Frank}\ \bibnamefont
  {Neese}},\ }\bibfield  {title} {\enquote {\bibinfo {title} {{The ORCA program
  system}},}\ }\href {\doibase 10.1002/wcms.81} {\bibfield  {journal} {\bibinfo
   {journal} {Wiley Interdiscip. Rev. Comput. Mol. Sci.}\ }\textbf {\bibinfo
  {volume} {2}},\ \bibinfo {pages} {73--78} (\bibinfo {year}
  {2012})}\BibitemShut {NoStop}%
\bibitem [{\citenamefont {Tesi}\ \emph {et~al.}(2016)\citenamefont {Tesi},
  \citenamefont {Lunghi}, \citenamefont {Atzori}, \citenamefont {Lucaccini},
  \citenamefont {Sorace}, \citenamefont {Totti},\ and\ \citenamefont
  {Sessoli}}]{Tesi2016}%
  \BibitemOpen
  \bibfield  {author} {\bibinfo {author} {\bibfnamefont {Lorenzo}\ \bibnamefont
  {Tesi}}, \bibinfo {author} {\bibfnamefont {Alessandro}\ \bibnamefont
  {Lunghi}}, \bibinfo {author} {\bibfnamefont {Matteo}\ \bibnamefont {Atzori}},
  \bibinfo {author} {\bibfnamefont {Eva}\ \bibnamefont {Lucaccini}}, \bibinfo
  {author} {\bibfnamefont {Lorenzo}\ \bibnamefont {Sorace}}, \bibinfo {author}
  {\bibfnamefont {Federico}\ \bibnamefont {Totti}}, \ and\ \bibinfo {author}
  {\bibfnamefont {Roberta}\ \bibnamefont {Sessoli}},\ }\bibfield  {title}
  {\enquote {\bibinfo {title} {{Giant spin-phonon bottleneck effects in
  evaporable vanadyl-based molecules with long spin coherence}},}\ }\href
  {\doibase 10.1039/C6DT02559E} {\bibfield  {journal} {\bibinfo  {journal}
  {Dalt. Trans.}\ }\textbf {\bibinfo {volume} {45}},\ \bibinfo {pages}
  {16635--16645} (\bibinfo {year} {2016})}\BibitemShut {NoStop}%
\bibitem [{\citenamefont {Atzori}\ \emph
  {et~al.}(2016{\natexlab{b}})\citenamefont {Atzori}, \citenamefont {Tesi},
  \citenamefont {Morra}, \citenamefont {Chiesa}, \citenamefont {Sorace},\ and\
  \citenamefont {Sessoli}}]{Atzori2016}%
  \BibitemOpen
  \bibfield  {author} {\bibinfo {author} {\bibfnamefont {Matteo}\ \bibnamefont
  {Atzori}}, \bibinfo {author} {\bibfnamefont {Lorenzo}\ \bibnamefont {Tesi}},
  \bibinfo {author} {\bibfnamefont {Elena}\ \bibnamefont {Morra}}, \bibinfo
  {author} {\bibfnamefont {Mario}\ \bibnamefont {Chiesa}}, \bibinfo {author}
  {\bibfnamefont {Lorenzo}\ \bibnamefont {Sorace}}, \ and\ \bibinfo {author}
  {\bibfnamefont {Roberta}\ \bibnamefont {Sessoli}},\ }\bibfield  {title}
  {\enquote {\bibinfo {title} {{Room-Temperature Quantum Coherence and Rabi
  Oscillations in Vanadyl Phthalocyanine: Toward Multifunctional Molecular Spin
  Qubits}},}\ }\href {\doibase 10.1021/jacs.5b13408} {\bibfield  {journal}
  {\bibinfo  {journal} {J. Am. Chem. Soc.}\ }\textbf {\bibinfo {volume}
  {138}},\ \bibinfo {pages} {2154--2157} (\bibinfo {year}
  {2016}{\natexlab{b}})}\BibitemShut {NoStop}%
\bibitem [{\citenamefont {Brown}(1973)}]{Brown1973}%
  \BibitemOpen
  \bibfield  {author} {\bibinfo {author} {\bibfnamefont {I.~M.}\ \bibnamefont
  {Brown}},\ }\bibfield  {title} {\enquote {\bibinfo {title} {{Concentration
  dependent relaxation times in organic radical solids}},}\ }\href {\doibase
  10.1063/1.1678980} {\bibfield  {journal} {\bibinfo  {journal} {J. Chem.
  Phys.}\ }\textbf {\bibinfo {volume} {4242}},\ \bibinfo {pages} {4242--4250}
  (\bibinfo {year} {1973})}\BibitemShut {NoStop}%
\bibitem [{\citenamefont {Bader}\ \emph {et~al.}(2014)\citenamefont {Bader},
  \citenamefont {Dengler}, \citenamefont {Lenz}, \citenamefont {Endeward},
  \citenamefont {Jiang}, \citenamefont {Neugebauer},\ and\ \citenamefont {{Van
  Slageren}}}]{Bader2014}%
  \BibitemOpen
  \bibfield  {author} {\bibinfo {author} {\bibfnamefont {Katharina}\
  \bibnamefont {Bader}}, \bibinfo {author} {\bibfnamefont {Dominik}\
  \bibnamefont {Dengler}}, \bibinfo {author} {\bibfnamefont {Samuel}\
  \bibnamefont {Lenz}}, \bibinfo {author} {\bibfnamefont {Burkhard}\
  \bibnamefont {Endeward}}, \bibinfo {author} {\bibfnamefont {Shang~Da}\
  \bibnamefont {Jiang}}, \bibinfo {author} {\bibfnamefont {Petr}\ \bibnamefont
  {Neugebauer}}, \ and\ \bibinfo {author} {\bibfnamefont {Joris}\ \bibnamefont
  {{Van Slageren}}},\ }\bibfield  {title} {\enquote {\bibinfo {title} {{Room
  temperature quantum coherence in a potential molecular qubit}},}\ }\href
  {\doibase 10.1038/ncomms6304} {\bibfield  {journal} {\bibinfo  {journal}
  {Nat. Commun.}\ }\textbf {\bibinfo {volume} {5}},\ \bibinfo {pages} {1--5}
  (\bibinfo {year} {2014})}\BibitemShut {NoStop}%
\bibitem [{\citenamefont {Atzori}\ \emph {et~al.}(2017)\citenamefont {Atzori},
  \citenamefont {Tesi}, \citenamefont {Benci}, \citenamefont {Lunghi},
  \citenamefont {Righini}, \citenamefont {Taschin}, \citenamefont {Torre},
  \citenamefont {Sorace},\ and\ \citenamefont {Sessoli}}]{Atzori2017}%
  \BibitemOpen
  \bibfield  {author} {\bibinfo {author} {\bibfnamefont {Matteo}\ \bibnamefont
  {Atzori}}, \bibinfo {author} {\bibfnamefont {Lorenzo}\ \bibnamefont {Tesi}},
  \bibinfo {author} {\bibfnamefont {Stefano}\ \bibnamefont {Benci}}, \bibinfo
  {author} {\bibfnamefont {Alessandro}\ \bibnamefont {Lunghi}}, \bibinfo
  {author} {\bibfnamefont {Roberto}\ \bibnamefont {Righini}}, \bibinfo {author}
  {\bibfnamefont {Andrea}\ \bibnamefont {Taschin}}, \bibinfo {author}
  {\bibfnamefont {Renato}\ \bibnamefont {Torre}}, \bibinfo {author}
  {\bibfnamefont {Lorenzo}\ \bibnamefont {Sorace}}, \ and\ \bibinfo {author}
  {\bibfnamefont {Roberta}\ \bibnamefont {Sessoli}},\ }\bibfield  {title}
  {\enquote {\bibinfo {title} {{Spin Dynamics and Low Energy Vibrations:
  Insights from Vanadyl-Based Potential Molecular Qubits}},}\ }\href {\doibase
  10.1021/jacs.7b01266} {\bibfield  {journal} {\bibinfo  {journal} {J. Am.
  Chem. Soc.}\ }\textbf {\bibinfo {volume} {139}},\ \bibinfo {pages}
  {4338--4341} (\bibinfo {year} {2017})}\BibitemShut {NoStop}%
\bibitem [{\citenamefont {Lunghi}\ \emph
  {et~al.}(2017{\natexlab{a}})\citenamefont {Lunghi}, \citenamefont {Totti},
  \citenamefont {Sessoli},\ and\ \citenamefont {Sanvito}}]{Lunghi2017}%
  \BibitemOpen
  \bibfield  {author} {\bibinfo {author} {\bibfnamefont {Alessandro}\
  \bibnamefont {Lunghi}}, \bibinfo {author} {\bibfnamefont {Federico}\
  \bibnamefont {Totti}}, \bibinfo {author} {\bibfnamefont {Roberta}\
  \bibnamefont {Sessoli}}, \ and\ \bibinfo {author} {\bibfnamefont {Stefano}\
  \bibnamefont {Sanvito}},\ }\bibfield  {title} {\enquote {\bibinfo {title}
  {{The role of anharmonic phonons in under-barrier spin relaxation of single
  molecule magnets}},}\ }\href {\doibase 10.1038/ncomms14620} {\bibfield
  {journal} {\bibinfo  {journal} {Nat. Commun.}\ }\textbf {\bibinfo {volume}
  {8}},\ \bibinfo {pages} {14620} (\bibinfo {year}
  {2017}{\natexlab{a}})}\BibitemShut {NoStop}%
\bibitem [{\citenamefont {Lunghi}\ \emph
  {et~al.}(2017{\natexlab{b}})\citenamefont {Lunghi}, \citenamefont {Totti},
  \citenamefont {Sanvito},\ and\ \citenamefont {Sessoli}}]{Lunghi2017a}%
  \BibitemOpen
  \bibfield  {author} {\bibinfo {author} {\bibfnamefont {Alessandro}\
  \bibnamefont {Lunghi}}, \bibinfo {author} {\bibfnamefont {Federico}\
  \bibnamefont {Totti}}, \bibinfo {author} {\bibfnamefont {Stefano}\
  \bibnamefont {Sanvito}}, \ and\ \bibinfo {author} {\bibfnamefont {Roberta}\
  \bibnamefont {Sessoli}},\ }\bibfield  {title} {\enquote {\bibinfo {title}
  {{Intra-molecular origin of the spin-phonon coupling in slow-relaxing
  molecular magnets}},}\ }\href {\doibase 10.1039/c7sc02832f} {\bibfield
  {journal} {\bibinfo  {journal} {Chem. Sci.}\ }\textbf {\bibinfo {volume}
  {8}},\ \bibinfo {pages} {6051--6059} (\bibinfo {year}
  {2017}{\natexlab{b}})}\BibitemShut {NoStop}%
\end{thebibliography}
%

\end{document}